# A novel method to separate circadian from non-circadian masking effects in order to enhance daily circadian timing and amplitude estimation from core body temperature

Phuc D Nguyen[1], Claire Dunbar[1], Hannah Scott[1], Bastien Lechat[1], Jack Manners[1], Gorica Micic[1], Nicole Lovato[1], Amy C Reynolds[1], Leon Lack[1], Robert Adams[1], Danny Eckert[1], Andrew Vakulin[1]*, Peter G Catcheside[1]*

1. Flinders Health and Medical Research Institute: Sleep Health, College of Medicine and Public Health, Flinders University, Bedford Park, South Australia 5042.

*. Co-senior authors with equivalent overall contribution to study and analysis oversight

Corresponding author: Phuc D. Nguyen

**Email:** ducphuc.nguyen@flinders.edu.au

**Author Contributions:** PN and PC conceived of the presented idea. PN performed the computations. PC verified the analytical methods. PC and AV supervised the findings of this work. All authors discussed the results and contributed to the final manuscript.

**Competing Interest Statement:** The authors have no conflicts of interest to declare. All co-authors have seen and agree with the contents of the manuscript and there is no financial interest to report.

**Classification:** Biological Sciences; Physiology.

**This PDF file includes:**

    Main Text
    Figures 1 to 2

**Abstract**

Circadian disruption contributes to adverse effects on sleep, performance, and health. One accepted method to track continuous daily changes in circadian timing is to measure core body temperature (CBT), and establish daily, circadian-related CBT minimum time (Tmin). This method typically applies cosine-model fits to measured CBT data, which may not adequately account for substantial wake metabolic activity and sleep effects on CBT that confound and mask circadian effects, and thus estimates of the circadian-related Tmin. This study introduced a novel physiology-grounded analytic approach to separate circadian from non-circadian effects on CBT, which we compared against traditional cosine-based methods. The dataset comprised 33 healthy participants (mean±SD 32±13 years) attending a 39-hour in-laboratory study with an initial overnight sleep followed by an extended wake period. CBT data were collected at 30-second intervals via ingestible capsules. Our design captured CBT during both the baseline sleep period and during extended wake period (without sleep) and allowed us to model the influence of circadian and non-circadian effects of sleep, wake, and activity on CBT using physiology-guided generalized additive models. Model fits and estimated Tmin inferred from extended wake without sleep were compared with traditional cosine-based models fits. Compared to the traditional cosine model, the new model exhibited superior fits to CBT (Pearson R 0.90 [95%CI; [0.83 - 0.96] versus 0.81 [0.55-0.93]). The difference between estimated vs measured circadian Tmin, derived from the day without sleep, was better fit with our method (0.2 [-0.5,0.3] hours) versus previous methods (1.4 [1.1 to 1.7] hours). This new method provides superior demasking of non-circadian influences compared to traditional cosine methods, including the removal of a sleep-related bias towards an earlier estimate of circadian Tmin.

**Significance Statement**

Circadian rhythm estimation from core body temperature (CBT) is challenging due to confounding influences of activity, sleep, and wake on current measurement approaches. This study introduces a novel physiology-based modelling approach to estimate and separate circadian from sleep, wake and activity effects on CBT, with demonstrably superior fits and estimates of the daily circadian-related CBT minimum time that are less biased towards sleep compared to traditional models. The method more effectively separates circadian and non-circadian effects in CBT data and holds major promise for enhanced tracking of daily circadian timing in real-world settings.

**Introduction**

Circadian rhythms are daily fluctuations in biological processes that follow an approximately 24-hour day-night cycle (1, 2). The central circadian pacemaker, located within the suprachiasmatic nucleus (SCN) in the brain, governs the body's overall circadian or 'body-clock' timing, which then signals to intracellular clock mechanisms throughout the body (3). These systems regulate many aspects of biology, including cellular metabolism, a key determinate of resting metabolic rate and core temperature, multiple organ functions, and strongly influence wake alertness and sleep propensity, timing and quality. Thus, circadian disruption, as occurs with shift-work, trans-meridian travel (i.e. jet-lag)(4) and circadian rhythm sleep-wake disorders (5). Hence, accurate individual level daily estimates of circadian phase would be of great value towards guiding targeted interventions to reduce the adverse impacts of circadian disruption or optimising circadian timing.

It is difficult to directly measure the timing of the central circadian pacemaker in humans (6), so research predominantly relies on markers of pacemaker activity and function. Core body temperature (CBT) minimum time (Tmin), plasma, salivary or urinary melatonin, and plasma cortisol are markers used to estimate the timing of the central circadian pacemaker, given well established phase relationships with sleep regulation (7–9). While dim-light melatonin onset assessment is typically regarded as the gold-standard, daily monitoring is costly and impractical given requirements for extended data collection under controlled conditions, including a delayed bedtime expected to impact sleep, next day functioning and circadian timing (10). Furthermore, sample analysis times and costs are typically too unfavourable to support daily circadian timing assessments and timely interventions. Consequently, while melatonin measurement may provide a more precise estimate of timing, application in real-world settings and rapid translation and utility in treatment is largely infeasible. Pragmatically, CBT is better suited to daily circadian timing estimation, particularly with the advent of accurate ingestible electronic capsule temperature sensors (11, 12).

Despite the practical utility of CBT measurement for circadian estimation, interpretation of CBT data is challenging because many factors simultaneously influence CBT. These include circadian influences in combination with variable effects of physical activity, sleep, metabolic rate, food intake, hydration level, and thermoregulatory behaviours (13) external factors such as radiant and convective heat gain or loss, ambient temperature, humidity, and clothing (14, 15). Appropriate modelling of CBT is therefore critically important to delineate the relatively small (±0.2 degree Celsius) endogenous circadian component from similar or larger magnitude masking effects of non-circadian, but inter-related, influences. This is particularly the case for sleep and activity which are known to have substantial effects on CBT and follow their own diurnal rhythm that may mask the CBT circadian rhythm (16, 17).

Theory-based parametric models, such as the traditional cosinor model (18), are amongst the simplest and most widely employed methods to estimate circadian timing from daily cyclical fluctuations in CBT. One or more harmonic components are often incorporated to help capture asymmetric components, including a "post-lunch dip", to improve model fits (19, 20). These parametric models are theory based and represent the main features of circadian rhythm fluctuations in CBT relatively well. However, this traditional approach fails to account for the multiple complex physiological aspects that influence CBT, such as sleep and exercise effects know to have a substantial impact on CBT. Therefore, the traditional approach remains best suited to CBT data collected in tightly controlled experimental settings (21) but has limited real-world utility through lack of flexibility needed to more accurately model asymmetric non-cosinor components of the circadian contribution to CBT (22). This introduces a significant translation hurdle into the real world, as control of sleep and activity is unrealistic and inherently problematic in uncontrolled real-world settings. This is particularly the case for sleep medicine, where effective and rapid measurement of circadian rhythm in patients with suspected circadian rhythm sleep-wake disorders is essential for informing advances in treatment choice, timing of treatment delivery (e.g. light, exogenous melatonin), and for monitoring efficacy in real-time.

Novel methods to better estimate the circadian component of CBT have been proposed, but none adequately account for known physiological confounding effects. Nonparametric models, such as spline regression (23), offer more flexibility than traditional cosine models without the need for specific assumptions regarding the shape of the circadian component of CBT. Other signal processing approaches, such as Fourier analysis (21), can also be used to decompose CBT data into sinusoidal waveforms of varying frequencies to help accommodate more complex circadian CBT components (24). A common approach to enhance model fit, and consequently, the reliability of endogenous circadian component timing and amplitude estimates, is to introduce one or more harmonics to the cosinor model. This improves model flexibility and fits, and yet the rationale for selecting any specific number of harmonics is not well grounded physiologically (23). Furthermore, most non-circadian influences on CBT are heavily influenced by human behaviours and other factors with substantial variability in timing and amplitude over each 24-hour cycle (13).

These masking or confounding effects on CBT pose a major challenge for deriving circadian effects from the CBT signal. Various time series linear demasking shapes (e.g., square wave or triangle) have been proposed to help remove sleep effects but are generally unreliable for accurate estimation of circadian phase (25). Furthermore, sleep and circadian effects are relatively small in amplitude compared to activity and especially exercise effects. Thus, more robust and more physiologically grounded analytical methods are required to help support daily estimation of endogenous circadian versus masking effects on CBT.

Here, we propose a novel physiologically-based analytical method, termed here as the recosinor model, to mitigate the masking effects known to be present in measured CBT data. This is intended to enhance the accuracy of circadian timing and amplitude estimates from CBT.

**Results**

We used data from an extended wakefulness protocol that examined the relationship between vestibular ocular measures, state sleepiness, and driving performance (ACTRN12621001610820). The dataset included 33 healthy participants (mean±SD 32±13 years old) with 39 hours of CBT data available for analysis (see Methods for details). The experimental design captured both sleep and extended wakefulness effects on CBT, allowing for a systematic evaluation of both circadian and sleep-wake contributions to CBT (Figure 1A), particularly useful for model development and testing against existing models. A generalized additive model (GAM) was used to formulate the overall model as a function of circadian, sleep-wake, and activity effects (Figure 1B) (see Methods). The shape of the intrinsic circadian rhythm was modelled using a recursive sine function which is a parametric model with similar overall behaviour to a cosine function, but with more flexibility regarding the shape and period of the function (see Methods). We also observed that the sleep-wake contribution to the measured CBT closely approximates a gamma distribution (Figure 1A), highly amenable to modelling with only two parameters for each component (Figure 1B). Furthermore, the integral of these gamma distributions approximate exponential decay functions hypothesized to underlie neurally driven homeostatic switching and cumulative decrement and recovery processes associated with sleep-wake transitions (9).

An example of model fitting and the estimation of endogenous circadian and homeostatic processes is shown in Figure 2A-C. This example highlights the capacity of the model to uncover and fit the main circadian-, sleep-wake-, and activity-dependent processes underlying the measured CBT (Figure 2B, C). In this example, the new model (recosinor) fit demonstrates a substantial improvement compared to the widely used cosinor and harmonic model (Figure 2A) (see Methods). Qualitatively, the estimated endogenous circadian curve (Figure 2B) closely resembles the shape and aligns well with the hypothesised circadian pattern (26). In contrast, the cosinor model, particularly with the first harmonic, tended to introduce additional and potentially non-physiological Tmin values and components of CBT. Furthermore, the recosinor model simultaneously estimates the contributions of sleep and wake effects on measured CBT, providing a separate assessment of the homeostatic sleep drive and wake contributions to CBT (Figure 2C). This additive model partitioning of circadian and sleep-related effects on CBT is a conceptual improvement for better understanding of the main underlying processes known to govern CBT.

At the group level, the recosinor model exhibited the highest Pearson R 0.900 ([95CI% 0.833 to 0.962], P <0.001) when compared to other models (Figure 2D). The widely used cosinor and first harmonic model displayed the second-best performance, although sometimes with unrealistic Tmin values (Figure 2A). Models such as Cosinor, Van der Pol, and recursive cosine function showed lower performance in comparison to both the new model (Recosinor) and the cosinor model including the first harmonic.

In comparison to the measured circadian Tmin from CBT determined from the second day without sleep, the estimated Tmin from the cosinor with the first harmonic model exhibited a systematic bias towards the initial hours of sleep (Figure 2D difference 1.4 [95%CI 1.1 to 1.7] hours, P < 0.001). In contrast, the new model showed no significant bias (difference of 0.2 [95%CI -0.5 to 0.3] hours, p = 0.082).

**Discussion**

Our new method provides superior circadian phase estimates compared to the traditional cosinor model. This was achieved by incorporating more flexible functions, but with minimal additional parameters needed to constrain model fits around the major known circadian, sleep-wake and activity influences on CBT. This novel approach, compared head-to-head with more traditional circadian modelling methods, showed superior model fits to CBT data collected over 2 consecutive days with and without sleep in an extended-wakefulness laboratory protocol. Our findings also suggest that the traditional cosinor method produces a biased estimate of the endogenous circadian phase due to CBT lowering effects of sleep. This new method is a promising improvement for providing real-time, reliable estimates of daily circadian timing. This method may be particularly useful for circadian rhythm assessments and tailoring circadian treatments for sleep disorders, where daily variation in both circadian and sleep timing are expected and problematic for estimation with traditional CBT models. While further validation of this new approach is required against gold-standard melatonin measurement and under constant routine protocols to compare model estimates against ground truth CBT rhythm, these initial findings are a promising advance toward improved CBT analysis, and circadian estimation in free-living humans.

These findings also underscore the importance of accounting for sleep-wake effects when estimating circadian timing (27). The influence of sleep-wake effects clearly introduces bias and thus inaccuracies in the estimation of both the phase, amplitude and period (Tau) of daily circadian temperature rhythms. Given a physiological modelling basis, there are also clinical translational implications for the underlying model parameter estimates and features such as circadian amplitude and Tmin which may be useful for sleep disorder diagnosis (28), including circadian misalignment. They may also usefully inform the choice and timing of effective treatments such as sleep re-timing, exogenous melatonin administration and light therapy, where daily circadian Tmin estimation is especially useful to determine the optimal timing of bright light exposure and avoidance (29).

A major advantage of our model approach is its simplicity and effectiveness, utilizing the GAM framework (30) that readily accommodates both linear and nonlinear covariates, along with flexible error term distributions in computationally efficient methods. This approach also substantially benefits from physiologically guided constraints informed by established knowledge regarding the temporal patterns of circadian and sleep-wake influences (22) and shorter time-scale dynamic effects of large activity and metabolic effects on CBT. Our approach aligns with and extends recent studies that have demonstrated promising outcomes by integrating cosinor analysis into a generalized linear model (31, 32), that also provides for adaptable selection of error distributions, resulting in improved model performance.

Several methodological limitations warrant consideration in future research. How closely our estimates of the circadian component of Tmin, inferred from extended wake without sleep, aligns with the "true" endogenous circadian Tmin is unknown. In future studies it will be important to compare Tmin estimates against independent measures of circadian timing via serial plasma or salivary melatonin analyses and under carefully-controlled conditions to minimize other effects to more appropriately observe Tmin in the raw data. Further studies are also needed to test for potential ultradian effects, which could be readily incorporated into the GAM model. However, given high model performance versus more traditional methods, future analyses should also consider that traditional harmonic components may only be useful to accommodate the superposition of asymmetrical sleep and circadian influences that could partly or potentially largely explain ultradian effects. Another limitation is that while our method performed well in young, healthy participants, validation in sleep disorder groups expected to show more variable sleep and circadian patterns remains to be investigated. Participant activity was also minimised in this study so further work is clearly needed to test and likely refine the modelling of activity and exercise effects. Exercise is expected to have larger, but shorter time-scale effects on CBT than circadian and sleep effects, but can be relatively easily accommodated with the addition of heart rate data as a proxy for metabolic rate as used in this study.

Despite some inevitable limitations, this new method lays the major theoretical and analytical foundations needed for more effective CBT data modelling and demasking of daily circadian timing from sleep and activity effects. More practical and effective demasking should be possible via integration of wearable sensor data validated against direct measures in relevant groups, enabling application of circadian estimation in several use-case settings such as treatment for circadian rhythm sleep-wake disorders.

**Materials and Methods**

**Data collection**

Participants attended the sleep laboratory and underwent a nine-hour sleep opportunity timed in accordance with their habitual nocturnal sleep timing period, before undergoing ~30 hours of extended wakefulness. Participants remained in the laboratory at all times under controlled dim lighting conditions (<5 lux) with limited activity and no opportunities to undertake intensive exercise. All participants received regular, fixed-time (every 3 hours) small calorie controlled cold meals (approximate 200 to 300 kcals), to mitigate meal ingestion and metabolic effects on CBT. CBT data were captured at 30-second intervals using ingested core body temperature recording capsules (BodyCap e-Celcius, Hérouville-Saint-Clair), which have demonstrated excellent validity and test-retest reliability (12).

**Statistical modelling of circadian, sleep and activity effects on CBT**

We used a generalized additive model (GAM) to formulate the overall model as a function of circadian, sleep and wake, and activity effects:

$$y_i = \beta_0 + f_c(t_i) + f_s(t_i) + f_a(hr_i) + \epsilon_i$$

This model describes $y_i$ as the measured CBT at time $t_i$, $\beta_0$ is the model baseline intercept, $f_c(t_i)$ is a function driven by the circadian process that influences CBT, $f_s(t_i)$ is a function to model the contribution of sleep-wake effects on CBT, $f_a(hr_i)$ is a heart-rate dependent function to account for activity dependent effects on metabolism and thermoregulation responses that influence CBT, and $\epsilon_i$ is an error term, which can be set to different distributions, for which we used the gamma distribution in this study to help better account for non-symmetrical temporal distributions compared to Gaussian distribution based models. We hereafter refer to this model as the "recosinor" model given similarities to the traditional cosinor model but with a more flexible recursive function for the main circadian component.

The shape of the intrinsic circadian rhythm was modelled using a recursive sine function which is a parametric based model with similar overall behaviour to a cosine function, but with more flexibility regarding to the shape and period of the function:

$$f_c(t_i) = \sin\left(2\pi \int_0^{t_i} \frac{1}{\tau_c(t_i)} dt + \frac{f_c(t_{i-1})}{k} + \theta\right)$$

Here, $\tau_c(t_i)$ is the time-dependent circadian period, $k$ is a shape factor and $\theta$ is the phase angle. When $k$ is large and $\tau_c$ is constant, the function $f_c(t_i)$ becomes a typical cosine function. However, a recursive function with these parameters is more flexible, and can be readily adjusted to accommodate the asymmetrical shape of circadian effects on CBT (22) without any need for harmonics, for which overfitting and artefactual additional daily peaks and troughs can be problematic with real-world data.

We observed that the sleep-wake contribution to the measured CBT follows a gamma distribution shape, which can be controlled by two parameters $\alpha$ and $\beta$ as follows;

$$f_s(t_i) = \frac{1}{\beta^{\alpha}\Gamma(\alpha)} t_i^{(\alpha-1)} e^{-\frac{t}{\beta}}$$

The integral of these gamma functions approximates an exponential decay function hypothesized to underlie neurally driven homeostatic switching and cumulative decrement and recovery processes associated with sleep-wake transitions (9). In our study, we used lights out and on times to approximate sleep-wake timing.

Activity and thermoregulation amongst other factors, such as food intake, clearly have transient effects on CBT that will impact model fits when not considered in CBT models. In real-world studies, determining food intake and activity levels and timing, and their effects on CBT is challenging. Thus, we chose a simple model of activity and thermogenic effects of food intake using a simple heart rate dependent term as a proxy for metabolic and thermoregulatory demands on the cardiovascular system to approximate their generalized effects CBT. Although simplistic, this approach is preferrable to ignoring activity effects on CBT, and heart rate is practical measurement choice that is readily captured via many existing wearable devices. The nonlinear relationship between heart rate and CBT is readily accounted for in the GAM model via a smoothing function. The model parameters are determined by optimising the mean absolute error between the fitted and measured CBT.

**Traditional models**

We compared the performance of our new model with several existing models, including cosinor, cosinor with the first harmonic component, and the Van der Pol oscillator model (33). The cosinor model was that of (18):

$$y_i = \beta_0 + \sum_{r=1}^{d} C_r \cos\left(\frac{2\pi r}{\tau_c} t_i + \theta_r\right)$$

where $C_r$ is the model coefficient. The subscript $r$ indicates the harmonics and $d$ is the number of harmonics. In this equation, we tested the cosinor without harmonics ($d = 1$) and cosinor with the first harmonic ($d = 2$).

The Van der Pol oscillator is described by the following ordinary differential equation (33):

$$\frac{d^2y}{dt^2} + \epsilon \frac{2\pi}{\tau_c}\left(1 - \frac{4}{\gamma^2}y^2\right)\frac{dy}{dt} + \left(\frac{2\pi}{\tau_c}\right)^2 y = 0$$

Where $y$ is circadian signal, $\epsilon$ and $\gamma$ are model parameters that control the shape and amplitude of the circadian rhythm.

**Data cleaning and measured Tmin**

CBT data for all models were initially cleaned to remove major non-physiological outliers associated with capsule ingestion, loss, equilibration within the gastro-intestinal tract and hot and cold fluid ingestion. Specifically, if the change in CBT exceeded 0.4 degrees per minute, all data points within a 10-minute window before and after this duration were removed. Additionally, data points exceeding 39 degrees or falling below 35 degrees were also removed.

The same individual data were then fitted to each model to evaluate model fits and Tmin. We defined "measured Tmin" as the clock time corresponding to the minimum CBT observed during the second night without sleep. This "measured Tmin" served as the primary estimate for the endogenous circadian component of Tmin on the first day of the protocol, which also included sleep influences on CBT and thus Tmin.

**Statistical analysis**

Pearson correlation analysis was used to quantify the correlation between estimated and measured CBT. Two-sample t-tests were employed to assess the bias in estimating Tmin from the new versus other models. Unless specified otherwise, the uncertainty of a point estimate was quantified using 95% confidence intervals. A p-value less than 0.05 was considered statistically significant.

**Data and code availability**

Our new "recosinor" model was constructed and tested using R version 4.2.0 (R Core Team) and is available via https://github.com/ducphucnguyen/recosinor. This package also includes example data. Additionally, we analysed other models, such as "cosinor," using the "cosinor" package (https://github.com/sachsmc/cosinor). All data used in this analysis will be publicly available.

**Acknowledgments**

The authors acknowledge the Commonwealth Department of Infrastructure, Transport, Regional Development and Communications, Office of Road Safety (ORS) grant Road Safety Innovation Fund –96 (RSIF2-96) for their financial funding of this research study. DJE is funded by a National Health and Medical Research Council (NHMRC) of Australia Leadership Fellowship (1116942).

**Figures and Tables**

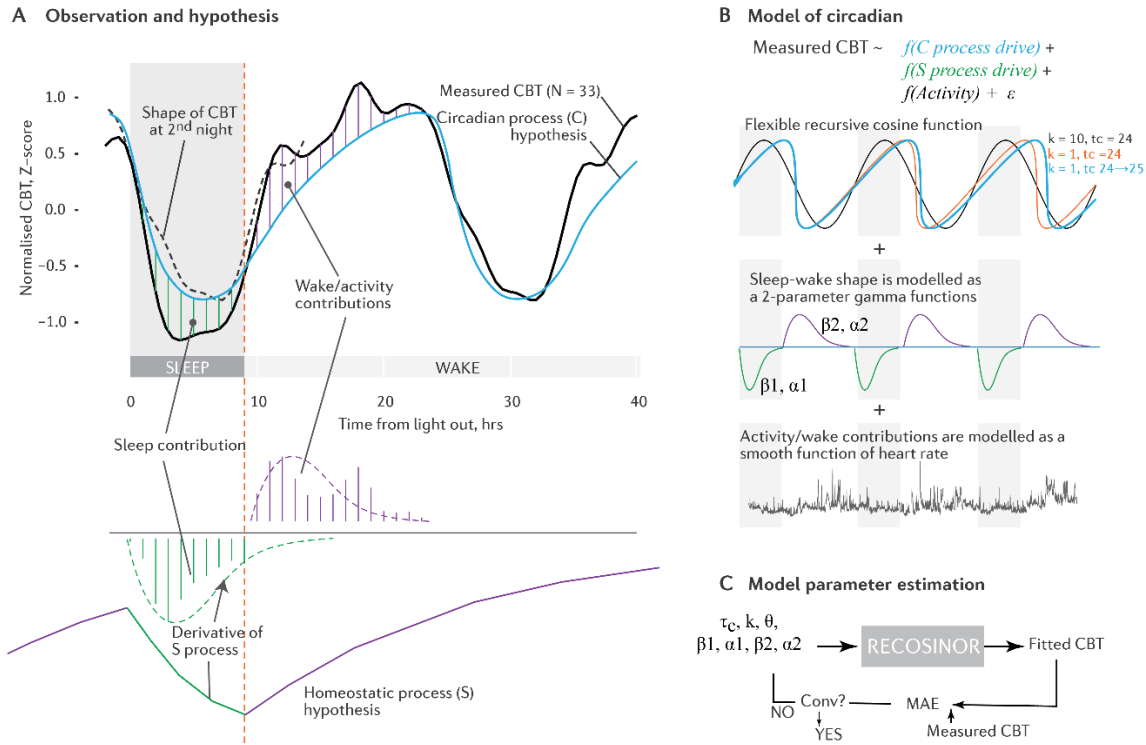

**Figure 1. Observations and model concept.  A**. Ensemble average core body temperature measurements from 33 participants aligned with the onset of lights-out. The hypothesised circadian process (process C depicted in blue) extracted from (26) exhibits a distinct shape that quite closely tracks the measured core body temperature (CBT), but with clear further influences from sleep and wake/activity effects (hatched areas). The gamma function (bottom panels) closely approximates the shape of these two important additional effects. **B**. The model is constructed in a GAM (Generalized Additive Model) format, employing a recursive sine function to capture the inherent circadian shape. **C**. Model parameters are determined through optimisation, aligning the fitted model with measured CBT data.

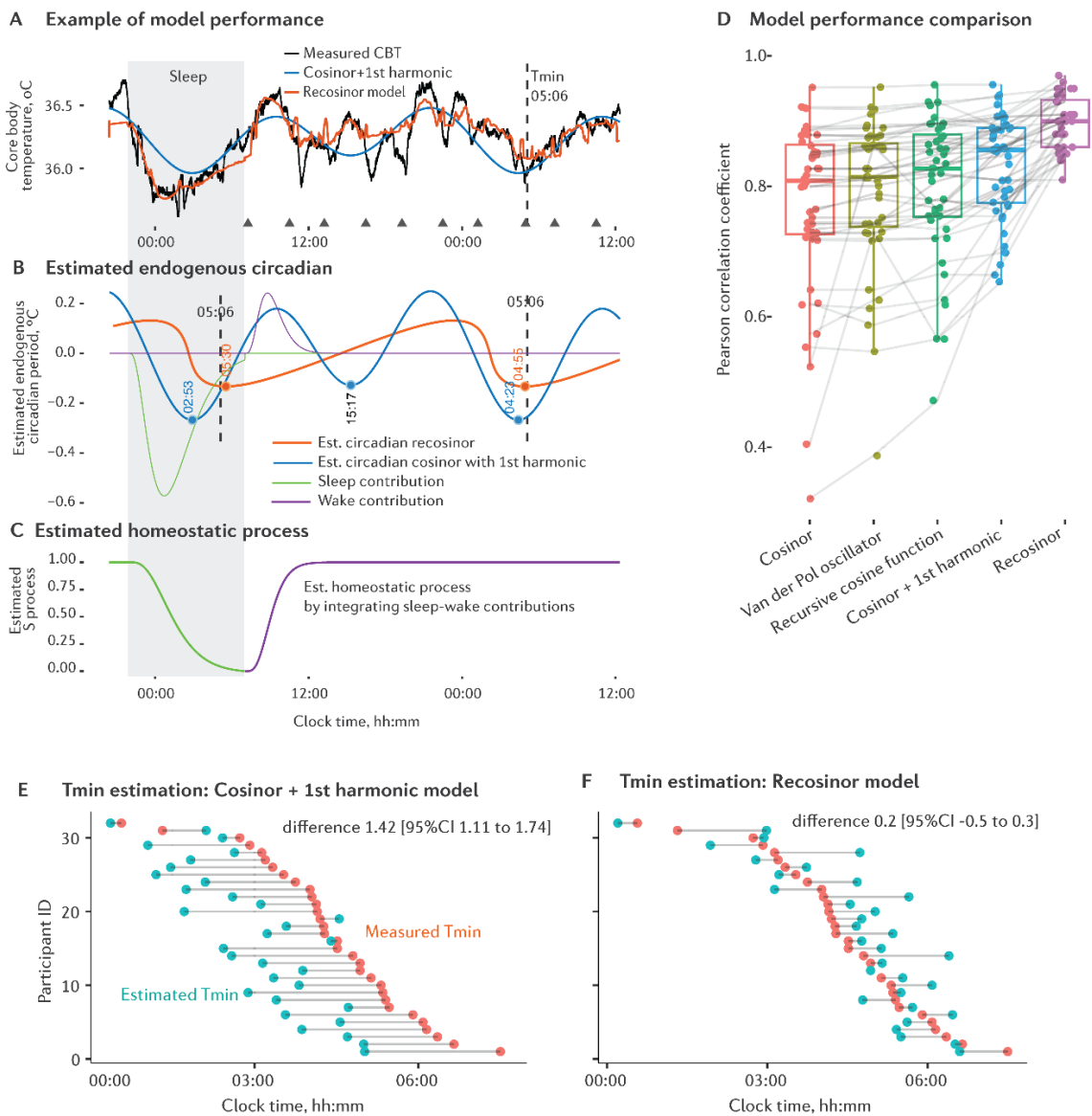

**Figure 2. Model hypothesis and performance. A**, Example of model fit between the recosinor model and cosinor + 1[st] harmonic model. Dashed line represents the measured circadian Tmin, estimated between 22:00 and 9:00 on the second night. Given the controlled low light and activity conditions and absence of sleep during extended wake, this measured Tmin should more closely reflect the endogenous circadian minimum point compared to the first night including sleep effects on CBT. **B**, Estimated endogenous circadian and sleep effects on CBT. **C**, Integrating sleep effects to estimate homeostatic sleep drive. **D**, Comparison of the correlation between the fitted CBT and the measured CBT with each model. **E, F**, Estimated Tmin during the first night with sleep versus the measured Tmin (see Methods for details), with participants ranked from latest to earliest measured Tmin.